\documentclass[preprint,floats,twocolumn]{aastex63}
\usepackage[utf8]{inputenc}
\usepackage[T2A]{fontenc}
\usepackage[russian,english]{babel}
\usepackage{amssymb}
\usepackage{amsmath}

\usepackage{multirow}

\usepackage{afterpage}
\usepackage{float}
\usepackage{placeins}

\begin{document}

\title[]{Electromagnetic Emission Produced by Three-wave Interactions in a Plasma with Continiously Injected Counterstreaming Electron Beams}

 \author[0000-0002-5577-8595]{V. V. Annenkov}
  \affiliation{Budker Institute of Nuclear Physics SB RAS, 630090, Novosibirsk, Russia}
 \affiliation{Novosibirsk State University, 630090, Novosibirsk, Russia}
 
 \author[0000-0002-8520-3207]{E. P. Volchok}
 \affiliation{Budker Institute of Nuclear Physics SB RAS, 630090, Novosibirsk, Russia}
 
  \author[0000-0001-5083-9777]{I. V. Timofeev}
 \affiliation{Budker Institute of Nuclear Physics SB RAS, 630090, Novosibirsk, Russia}

\keywords{plasmas -- radiation mechanisms: non-thermal -- Sun: radio radiation}


 \begin{abstract}
Three-wave interactions between Langmuir and electromagnetic waves in plasma with unstable electron flows are believed to be the main cause for type II and III solar radio emissions. The narrow band of type II bursts requires to assume that this radiation is generated in some local regions of shock fronts traveling in the solar corona, where the specific conditions for the enhancement of electromagnetic emissions near the plasma frequency harmonics are created. The reason for such enhancement at the second harmonic may be the formation of counter-streaming electron beams. There are different opinions in literature on whether the second harmonic electromagnetic emission in the presence of an additional beam can be efficient enough to markedly dominate emissions produced by a single beam. In the present paper, we carry out particle-in-cell simulations of the collision of two symmetric electron beams in plasma with open boundary conditions and show that the efficiency of beam-to-radiation power conversion can be significantly increased compared to models with periodic boundary conditions and reach the level of a few percent if three-wave interactions with electromagnetic waves near the second harmonic of the plasma frequency becomes available for the most unstable oblique beam driven modes.

 \end{abstract}

\section{Introduction}

Electromagnetic emission in the radio range is one of the main sources of information about the processes occurring both near the solar surface and in the heliosphere. Electron beams generated during magnetic reconnection \citep{Chen2018}, solar flares \citep{Kupriyanova2020} and coronal mass ejections (CME) \citep{Petrosian2016} provide free energy for a wide variety of different radio emissions such as solar radio bursts of Type II \citep{Roberts1959,Nelson1985,Gopalswamy2018,Kahler2019}, Type III \citep{Wild1963,Gurnett1976,Robinson1998,Thurgood2015,Che,Krafft2020}, Type IV \citep{Boischot1958,Vasanth2019} and bursts with a zebra structure \citep{Zheleznyakov2012,Kuznetsov2013}. Starting with the pioneering work of  \citet{Ginzburg1958},  a large number of radiation mechanisms based on different linear and nonlinear wave processes in plasma has been considered. A detailed description of these mechanisms can be found in previously cited papers, as well as in  the monograph \cite{Robinson2000}, textbook \cite{Aschwanden2006} and  review \cite{Reid2014}. Electron beam-plasma interaction is also considered as the possible cause of sub-THz solar flares \citep{Sakai2006,Zaitsev2013,Kontar2018} which became available for detection only recently \citep{Kaufmann2004,Kaufmann2009}. 

In laboratory conditions, similar processes of electromagnetic emission from a plasma under the injection of weakly relativistic electron beams (from 100 keV to 1 MeV) are presently studied in experiments at the GOL-PET facility (BINP SB RAS) \citep{Ivanov2015,Arzhannikov2016,Arzhannikov2020}. Although the main goal of these experiments is the search for efficient regimes of beam-to-radiation power conversion suitable for creation of a high-power THz source, the GOL-PET facility is also used as a test bed for proving fundamental theoretical insights on the physics of electromagnetic emission from a turbulent beam-plasma system \citep {Timofeev2012c,Timofeev2015,Annenkov2016, Annenkov2019}. Thus, the same physical processes are now actively studied by two scientific schools focused on different fields:  solar radio emissions and efficient THz generation in laboratory. Although the corresponding regimes of beam-plasma interaction differ in the density of beams and the presence of a guiding magnetic field, the general concepts of the mechanisms responsible for conversion of the beam kinetic energy into radiation energy in different approaches overlap significantly. In particular, recently there have appeared works \citep{Ganse2012b,Ganse2012,Ganse2014,Ziebell2015,Thurgood2015,Thurgood2016,Henri2019,Lee2019} on the numerical simulation of emission processes in a beam-plasma system, which were motivated by the problems of solar radio flares, but in terms of the parameters available for calculating, they were limited to the regimes of sufficiently dense beams typical to laboratory experiments. In this regard, in our opinion, there is a need to compare theoretical insights on the mechanisms of radiation generation in the beam-plasma system, which have been recently obtained by various scientific schools.

The main difference between the theoretical description of the beam-plasma interaction in a laboratory experiment and the description of similar processes in solar plasma is the spatial limitation of the system and the need to set realistic boundary conditions. Indeed, it is impossible to imagine a situation where a nonequilibrium distribution of beam electrons in an experiment is uniformly created at once in the whole plasma volume. In order to reach the plasma regions far from the injection site, the beam must travel a certain distance in it. But since its distribution is unstable, it must lose a part of its energy to excite plasma oscillations on its way. Thus, if the source of accelerated electrons is localized in space, then the collective relaxation of such a beam cannot be correctly reproduced in spatially homogeneous models with periodic boundary conditions. Nevertheless, such models are very popular in studies of solar radio bursts. The main drawback of the so-called temporal problem, when the beam instability develops simultaneously in the entire space, is the finite margin of the beam nonequilibrium.
The beam pumps the wave only until the depth of its potential well in the accompanying reference frame is sufficient to trap beam electrons. After this, the evolution of both waves resonant with the beam and waves from the rest of the turbulent spectrum proceeds in the absence of any energy inflow. In contrast, in the problem of beam propagation from a local source, the excited plasma oscillations, having a very low group velocity, accumulate at a certain distance from the injection site and form a spatially localized wave packet. The oscillations in this packet are pumped continuously by "fresh" electrons coming from the source, and their amplitude is saturated at a much higher level than in the temporal problem \citep{Timofeev2006,Timofeev2010}. The high local energy density of plasma oscillations leads to the fact that the beam relaxation and the associated processes of EM emission proceed along a different path than the scenario of weak turbulence predicts, even for beams with a low relative density $\sim 10^{-3}$ \citep{Timofeev2010,Annenkov2019}. In this paper, we will consider how the transition to a more realistic problem of beam injection affects the efficiency of radiation generation at the second harmonic of the plasma frequency in a system of colliding electron beams.

Emission processes in non-magnetized plasma with colliding beams have recently been studied in relation to the issue of a narrow line-width of type II solar radio bursts \citep{Ganse2012b,Ganse2012,Thurgood2015,Ziebell2016}. The generally accepted picture of this phenomenon \citep{Reiner1998} suggests the drift acceleration of electrons in curved sections of the shock front which is initiated by coronal mass ejections (CME). The flow of accelerated electrons from this spatially localized source propagates in the foreshock region where it excites Langmuir waves $L$ due to the two-stream instability. It is believed that the subsequent processes of EM emission repeat the scenario of type III radio bursts \citep{Ginzburg1958}. By participating in three-wave interactions with ion-acoustic oscillations $S$, resonant Langmuir waves $L$ produce not only EM waves near the plasma frequency ($L\rightarrow S+T_{\omega_p}$), but also a population of secondary Langmuir waves $L'$ ($L\rightarrow S+L'$) that propagate against the direction of the beam. Subsequent coalescence of forward and backward Langmuir waves generate radiation at the second harmonic of the plasma frequency ($T_ {2\omega_p}$) \citep{Melrose1986}:
\begin{equation}\label{tw}
L+L'\rightarrow T_{2\omega_p}
\end{equation}
In a finite magnetic field, the description of similar processes requires to take into account the splitting of EM waves by polarizations (X and O modes) \citep{Willes1997}. The interpretation of EM emission based on the above-mentioned nonlinear processes has been used to explain both the data of laboratory beam-plasma experiments \citep{Whelan1981,Schneider1982,Timofeev2012c} and the results of PIC simulations of type II and III radio bursts \citep{Kasaba2001,Sakai2005,Umeda2010,Thurgood2015}. The most detailed and convincing confirmation of the existence of a weakly turbulent scenario of EM emission at plasma frequency harmonics has been obtained in recent works \citep{Henri2019,Lee2019}, when rather low beam densities $n_b/n_p\sim 10^{-3}$ became available for PIC models with periodic boundary conditions. A common feature of the aforementioned laboratory and numerical experiments on the interaction of a single electron beam with plasma was a very low efficiency of beam-to-radiation energy conversion which does not exceed several units of $10^{-5}$. To explain the narrow band of type II radio bursts, it is necessary to assume that the radiation is generated at the local section of the CME's shock front which should be somehow distinguished in comparison with other parts of the front characterized by different values of the plasma frequency. \cite{Ganse2012b,Ganse2012} have suggested that such a feature could be either the intersection of two shock fronts, or the close location of the bends of a front, between which counter streams of accelerated electrons are formed. The presence of the second beam can significantly increase the efficiency of the second harmonic EM emission. Indeed, in the case of a single beam, low efficiency of the second harmonic emission  is caused by the presence of an intermediate stage of electrostatic decay. It automatically means that the vast majority of the beam energy is wasted on excitation of a wide spectrum of non-radiating electrostatic oscillations and consequent plasma heating. The energy sink from the resonant to nonresonant part of wave spectum is reduced if both electrostatic waves participating in the three-wave process (\ref{tw}) are excited directly by two counterstreaming electron beams. The enhancement of harmonic radiation in such a system has been not only confirmed in numerical simulations, but also demonstrated experimentally \citep{Leung1981,Intrator1984,Schumacher1993}. \cite{Ziebell2016} have studied generation of EM waves by counter-streaming beams by solving equations of weak turbulence and found that its enhancement is not so pronounced compared to the single beam case as reported by \cite{Ganse2012}.

In this paper, we will show that there are two reasons why the efficiency of the second harmonic emission from a plasma with counter-streaming electron beams should be higher than in previous works \citep{Ganse2012b,Ganse2012,Ziebell2016}. The first reason is the fundamentally nonuniform character of beam-plasma interaction which is not accounted for in models with periodic boundary conditions. Really, continuous pumping of beam energy to spatially localized wavepackets leads to higher densities of Langmuir wave energy than in the temporal problem. This affects not only EM emission, but also ion dynamics which can no longer be described in terms of sound waves and is more governed by ponderomotive forces of high-frequency fields. The second reason is the possibility to tune the system parameters in such a way that the most unstable counterpropagating beam-driven modes begin to participate in the coalescence process (\ref{tw}). First, we have demonstrated this possibility \citep{Timofeev2014b} using PIC simulations with periodic boundary conditions imposed along the beam axis. By calculating the growth rate of oblique instabilities in the framework of the exact relativistic kinetic theory \citep{Timofeev2013}, we have found the regime when dominant beam-driven modes can merge into  transversely propagating EM waves and shown that the second harmonic emission in this regime is several times more efficient than in the case when the fastest growing oscillations drop out of three-wave interaction. The first attempt to demonstrate the performance of this regime in a more realistic model with continuous injection of beams through plasma boundaries \citep{Volchok2019a} has been failed, since high beam densities used there make the system unstable against filamentation and favor another radiation mechanism based on the head-on collision of electrostatic plasma waves with different potential profiles \citep{Timofeev2017b,Annenkov2018}.    

Thus, the aim of the present paper is twofold. On the one hand, we want to demonstrate the real efficiency of the second harmonic emission in a system of counter-streaming electron beams by using PIC simulations with open boundary conditions. Such simulations are best suited to study instabilities of beams produced  by spatially localized sources (for example, by local sites of CME's shock fronts in the problem of type II radio-bursts). The open-boundary problem does also allow us to determine what fraction of beam power is capable of being converted to the power of EM emission from the whole interaction region. Since this efficiency can reach the level of a few percent, this may be an argument in favor of the strong dominance of regions with double-hump distributions in producing the observed type II radio emission.  On the other hand, such an efficient radiation mechanism looks attractive for generating high-power THz radiation in laboratory conditions. For this reason, we search for the regime with the enhanced second harmonic EM emission at low relative beam densities, which is stable against compression of beams by their own magnetic fields, and propose the set of parameters at which laboratory experiments can prove the discussed effect.

In Section \ref{EffReg}, the condition of synchronism between unstable beam-driven oscillations and radiated electromagnetic waves is formulated, and the results of a search for an efficient radiation regime for parameters interesting for experimental implementation are presented (beam energy $\approx1$ MeV and density $n_b = 0.005n_0$, where $n_b$ is the beam density and $n_0$ is the plasma density). Section \ref{SimFrame} describes the computational code, simulation layout, as well as  beam and plasma parameters. In Section \ref {OneBeam}, we discuss the features of beam-plasma interaction during the injection of a single beam into a finite-size plasma and investigate the convergence of the simulation results with the number of particles in a cell. Section \ref{SimRes} presents the results of PIC simulation in the case of two beams. The efficient radiation regime is compared with close regimes in which the most unstable beam modes cease to satisfy the three-wave condition. The Section \ref{Conc} contains a conclusion.

\section{Three-wave interaction}\label{EffReg}

Let us consider generation of EM radiation during the coupling of oblique oscillations excited in a plasma by counterstreaming electron beams.

\subsection{Synchronism condition}
Let identical electron beams with the density $n_b$ propagate towards each other in an infinite homogeneous plasma with the density $n_0$ along longitudinal direction (the $x$ axis). Due to the two-stream instability, the beams excite counterpropagating plasma oscillations with the frequency $\omega_b$ close to the plasma frequency $\omega_p = \sqrt{4\pi e^2 n_0/m_e}$, where $e$ and $m_e$ are the charge and mass of an electron.  The symmetry of the system allows one to present the wave vectors of the most unstable modes as follows:
\begin{equation}\label{kvec}
\mathbf{k_1^l}=(k_\parallel,k_\perp,0),\qquad \mathbf{k_2^l}=(-k_\parallel,k_\perp,0).
\end{equation}
EM radiation occurs when the following conditions for the three-wave interaction of these beam-driven modes with a transversely propagating EM wave are satisfied:
 \begin{gather}\label{TREE}
    \begin{cases}
\mathbf{k_1^l}+\mathbf{k_2^l}=\mathbf{k_3^t},\\
\omega_b(\mathbf{k_1^l})+\omega_b(\mathbf{k_2^l})=\omega_t(\mathbf{k_3^t}),
   \end{cases}
\end{gather} 
where $\omega_t$ is the eigenfrequency of electromagnetic wave in plasma and $\mathbf{k_3^t}$ -- its wave vector.

For the system considered, it means that the most intense radiation will be directed across the beams propagation axis ($\mathbf{k_3^t} = (0,0,2k_\perp)$) and its frequency will be localized near the doubled frequency of the beam-driven mode
\begin{gather}\label{omt}
\omega_t=2\omega_b.
\end{gather} 

In a cold magnetized plasma, only the ordinary mode (O mode) is strictly electromagnetic. Its dispersion law has the same form as in an unmagnetized plasma:
\begin{gather}\label{omtd}\omega_t(2k_\perp)=\sqrt{1+4k_\perp^2}.\end{gather} 
(hereafter, we will measure all frequencies in units of the plasma frequency $\omega_p$ and wave vectors in units of $\omega_p/c$, where $c$ is the speed of light in vacuum). Waves with extraordinary polarization (X mode) have a finite potential component, but, for the parameters considered in the paper, the difference  of their dispersion in the vicinity of the doubled plasma frequency from the law (\ref {omtd}) can be neglected. Since the spectral line of the unstable wave $\omega = \omega_b + i\Gamma$ is broadened by its growth rate $\Gamma(k_\parallel, k_\perp)$, the resonance conditions of the three-wave interaction (\ref{TREE}) for both modes of a magnetized plasma must be satisfied with the following accuracy:
  \begin{equation}\label{cond}
\left|\omega_b\left(k_\parallel,k_\perp\right)-\sqrt{k_\perp^2+\frac{1}{4}} \right| \leqslant \Gamma\left(k_\parallel,k_\perp\right),
\end{equation} 
where $\omega_b\left(k_\parallel, k_\perp\right)$ is the real part of the frequency of the beam mode that is determined from the solution of the linear dispersion equation for the beam-plasma instability. The condition (\ref{cond}) defines a bounded region in the $k$-space. We will call the regime in which the global maximum of the growth rate falls in this region  as  \textit{efficient} one, since plasma oscillations with the maximal spectral energy density are involved in the generation of EM radiation in this case.

 \subsection{Realization of efficient regime}\label{LinEffReg}
 
\begin{figure*}[ht!]
	\begin{center}
		\includegraphics[width=1\linewidth]{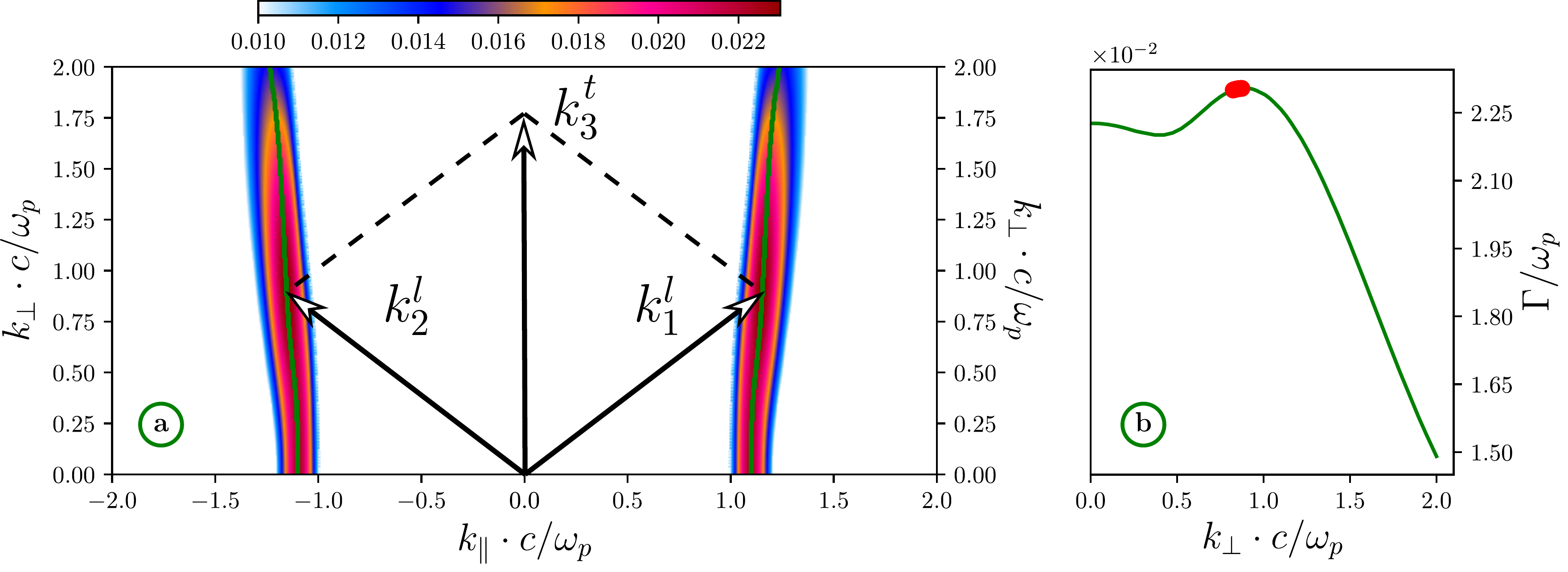}
		\caption{(a) The growth rate map for the beam-plasma instability $\Gamma\left(k_\parallel,k_\perp\right)$  in the efficient regime. The green line $k_\perp=k_\perp(k_\parallel)$ marks the maximal growth rate achieved for each $k_\perp$. (b) $\Gamma(k_\perp)$ along the green line (red points indicates the  region of the three-wave interaction).}
		\label{fig:EffInc}
	\end{center}
\end{figure*}
In \cite{Timofeev2014b}, we have found such an efficient regime for beams with the kinetic energy 1 MeV and the relative density $n_b=0.05$. The parameters have been optimized by varying the anisotropic temperature of the beams and the magnitude of the longitudinal external magnetic field $B=(B_0,0,0)$. The high efficiency of this regime has been confirmed using 2D3V particle-in-cell simulations with periodic boundary conditions.

The same regime has been later simulated in a model of a finite-size plasma with continuously injected beams \citep{Volchok2019a}. This model is more realistic, since it allows a constant inflow of new beam particles from a localized source. In addition, this model does not impose restrictions on the possible spectrum of electromagnetic oscillations in the system. These simulations have shown that the injection of beams with a sharp front creates a seed for the priority excitation of purely longitudinal oscillations, despite the fact that the linear theory predicts the excitation of oblique modes. A similar effect has been observed in \cite{Annenkov2019}. This fact, as well as the significant compression of dense beams by their own magnetic fields, led to the turning on the emission mechanism due to the head-on collision of longitudinal plasma oscillations with different transverse potential profiles \citep{Timofeev2017b,Annenkov2018}. 

Although the effect of dominant excitation of purely longitudinal waves can be completely excluded by the use of a smooth beam front with the typical growth time $\tau = 50\omega_p^{-1}$, we have failed to demonstrate the performance of the discussed three-wave scheme for the given beam density due to the rapid transition of instability to the strongly nonlinear stage. In regimes with lower beam densities, one should expect a more extended stage of dominance of oscillations growing with the maximal growth rate predicted by the linear theory. For this reason, in this work, we will search for the efficient regime at the lower beam density $n_b=0.005$.

We assume that the plasma and beam electrons have Maxwellian momentum distributions:
\begin{gather}
f^{(\varsigma)}\left(p_\perp,p_\parallel\right)\propto\exp\left(-\dfrac{p^2_{\perp}}{\Delta p^{(\varsigma)2}_{\perp}}-\dfrac{\left(p_{\parallel}-P^{(\varsigma)}\right)^2}{\Delta p^{(\varsigma)2}_{\parallel}}\right),
\end{gather}
where $\varsigma= (e, b)$ marks the plasma or the beam electrons, $P^{(\varsigma)}$ is the directed momentum, $p_\perp$ and $p_\parallel$ -- transverse and longitudinal with respect to the external magnetic field particle momenta. The temperature can be determined as $$T_{\parallel,\perp}=\dfrac{\Delta p^{(\varsigma)2}_{\parallel,\perp}}{2} m_e c^2.$$

The temperature of the background plasma of the order of tens of eV has a negligible effect on the instability growth rate \citep{Timofeev2013}. Thus, we fix it at the value $T_\parallel^{(e)}=T_\perp^{(e)}=T^{(e)}=80$ eV. The choice of this temperature is due to the necessity of  numerical scheme stability in further PIC simulations. For this purpose, the computational grid step must be $\Delta x <3.4\lambda_D$ \citep{hockney1988}, where $\lambda_D$ is the Debye length.

Relativistic velocities of beams  lead to the preferential growth of oblique unstable modes due to the mass anisotropy effect \citep{Bret2010a}. On the other hand, oblique instabilities are strongly affected by an external magnetic field and beam thermal spreads: an increase in the longitudinal temperature leads to a slow suppression of longitudinal instabilities, while the growth of transverse spreads -- to a rapid suppression of oblique ones.

We  search for the efficient regime in the multi-parameter space $(v_b,\Omega_e,T^{(b)}_\parallel,T^{(b)}_\perp)$, where $\Omega_e=eB_0/m_ec$ is the electron cyclotron frequency in the external magnetic field $B_0$. For this purpose, we use the numerical algorithm for solving the dispersion equation of a collisionless magnetized plasma in the framework of the exact relativistic kinetic theory \citep{Timofeev2013}. In solving the dispersion equation, we neglect the contribution of ions which do not significantly affect high-frequency oscillations. The parameters corresponding to the efficient regime are presented in the table \ref{table:EffParams}.

\begin{deluxetable}{ccc}
\tablecaption{Parameters of the efficient regime. \label{table:EffParams}}
\tablehead{\colhead{Parameter}&\colhead{Designation}&\colhead{Value}}
\startdata
Beams density & $n_b/n_0$	&  $0.005$\\
Beams velocity &$v_b/c$	&  $0.94$\\
Magnetic field &$\Omega_e/\omega_p$ & $0.29$\\
Plasma electrons temperature& $T^{(e)}$ & $80$ eV\\
\multirow{2}{*}{Beams electrons temperature} &$T^{(b)}_\parallel$ & $207.36$ keV\\
&$T^{(b)}_\perp$ & $25.0801$ keV\\
\enddata
\end{deluxetable}

The growth rate map $\Gamma(k_\perp,k_\parallel)$ for the beam-plasma instability in this regime is shown in Figure \ref{fig:EffInc} (a). Green curves show the position of the local growth rate maximized over $k_{\perp}$. The arrows $k^l_1$ and $k^l_2$ correspond to the wave vectors of the most unstable beam-plasma modes. The wave vector of radiated EM wave $k^t_3$ is the result of their summation. Figure \ref{fig:EffInc} (b) shows how the growth rate $\Gamma(k_\perp)$ changes along the green line. Red dots correspond to the values satisfying the condition (\ref{cond}).

The next step after the finding of optimal parameters is a numerical experiment using the particle-in-cell method.

\section{PIC model}\label{SimFrame}

This section describes a particle-in-cell model with open boundaries and discusses the simulation layout.

\subsection{Numerical schemes}

For numerical simulations, we use our own 2D3V Cartesian parallel code implemented for Nvidia GPGPU \citep{Lindholm2008}. It is based on standard computational schemes: the Yee solver of Maxwell equations for EM fields \citep{Yee1966}, the Boris \citep{Boris1970} scheme for solving the equation of motion for macro-particles with a parabolic form factor, and the charge-conserving Ezirkepov scheme \citep{Esirkepov2001} for calculations of currents.

\subsection{Open boundary conditions}
The main feature of this code is implementation of self-consistent open boundary conditions which allows us not only to realize continuous injection of particle beams through plasma boundaries, but also maintain a physically correct compensating current of plasma particles across the boundaries of the computational domain. For this purpose, the so-called plasma buffers located at the ends of the plasma column are used. The role of these buffers is to simulate the presence of real plasma behind the calculated region. Their goal is to delete the outgoung particles, to create particles that would have to pass through its boundary in the case when there was a real plasma behind them and also maintain the correct current values in the boundary cells of the simulated plasma.

A more detailed description of these open boundary conditions can be found in \citet{Annenkov2019a}. Correct implementation of these conditions is crucial for the problem of steady-state beam injection \citep{Berendeev2018}. Our code is able to simulate the injection of a beam into a plasma at the parameters of the GOL-3 experiments (BINP SB RAS) at realistic time-scales up to $t = 10^4\ \omega_p^{-1}$ \citep{Annenkov2019}. A good agreement between the results of numerical and laboratory experiments has been obtained.

\subsection{Simulation layout}
\begin{figure}[h]
	\includegraphics[width=0.92\linewidth]{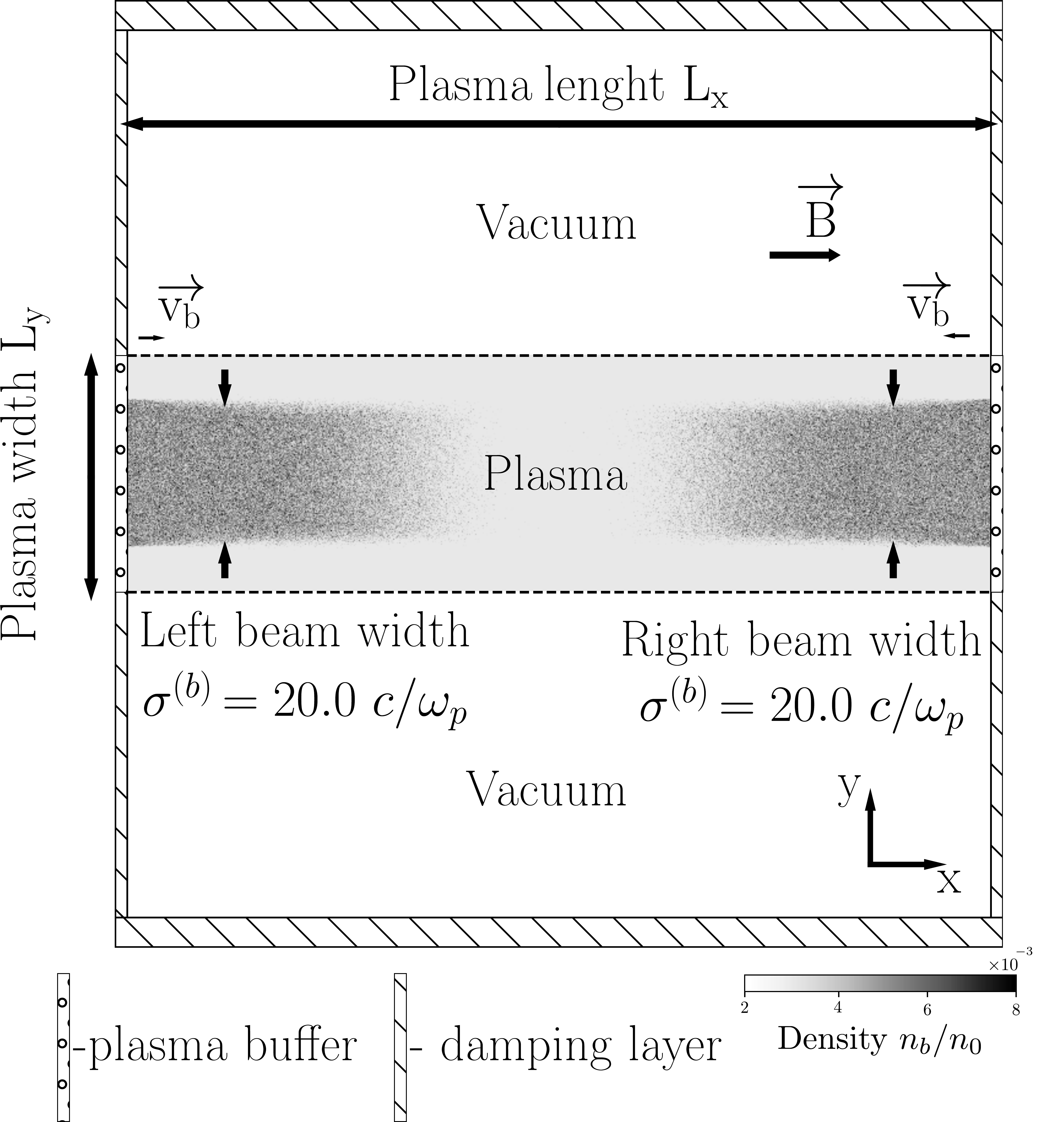}
	\caption{Simulation layout.}
	\label{fig:layout}
\end{figure}

Figure \ref{fig:layout} shows the layout of the simulation box used in this work. A plasma column is located at the centre of the system. Plasma buffers are located at the ends of the plasma slab. The width of the plasma equals to $L_y=32$~$c/\omega_p$. Plasma particles are initiated with no directional velocities. Through the plasma boundaries, we inject electron beams with the diameter $\sigma^{(b)}=20$~$c/\omega_p$. This size is large enough for the development of the oblique instability with the transverse wavelength  $\lambda_\perp=2\pi/k_\perp\approx2\pi/0.9\approx7$~$c/\omega_p$. A further increase in the width of the beam-plasma system is limited by available computing resources. In order to avoid creating a seed for the development of the longitudinal two-stream instability, we use a smooth beam front with the rise-time $\tau=50\omega_p^{-1}$. Plasma column is separated in the transverse direction from the boundaries by vacuum layers with the size $32$~$c/\omega_p$. The entire system is immersed in an uniform longitudinal magnetic field $B_x$. We set the magnitude of the magnetic field through the ratio of the electron cyclotron frequency $\Omega_e$ to the plasma frequency $\omega_p$.  The simulated region is surrounded by boundary layers absorbing EM radiation. Their description can be found in \cite{Annenkov2019a}. The spatial grid step is the same in all directions and equal to $\Delta x=\Delta y=0.04c/\omega_p$, time step is $\Delta t=0.02\omega_p^{-1}$.

\section{Single beam injection}\label{OneBeam}
In the case of periodic boundary conditions, the two-stream instability developes along the entire length of the simulated system, that is why one can limit the longitudinal plasma size in such a uniform system by several wavelengths. In the problem of steady-state beam injection, the instability developes in space, which requires much longer distances and times. 
The nonlinear stage of beam trapping in such a problem is accompanied by the formation of a spatially compact wave packet that is located at a relatively large distance from the injection point, $\sim v_b/\Gamma$ \citep{Sigov1996,Umeda2002,Timofeev2010,Sakai2005}. It is the region where the largest amplitude of plasma oscillations will be observed. Therefore, when we study emission processes by counterstreaming beams, it is necessary to provide the best overlapping of  relaxation regions for each beam. This can be achieved by choosing an appropriate length of the system in simulations of a single beam.

It is worth noting that the scenario of the beam instability development is unique for each specific implementation of the distribution function of the beam and plasma particles. Therefore, two simulations with the same geometry and macroscopic parameters (temperature, beam energy, etc.) will not be completely identical. For this reason, it is also necessary to assess how large the scatter between two shots with the same initial parameters is and to study convergence depending on the number of particles in a cell.

\begin{figure}[h!]
\centering
	\includegraphics[width=1\linewidth]{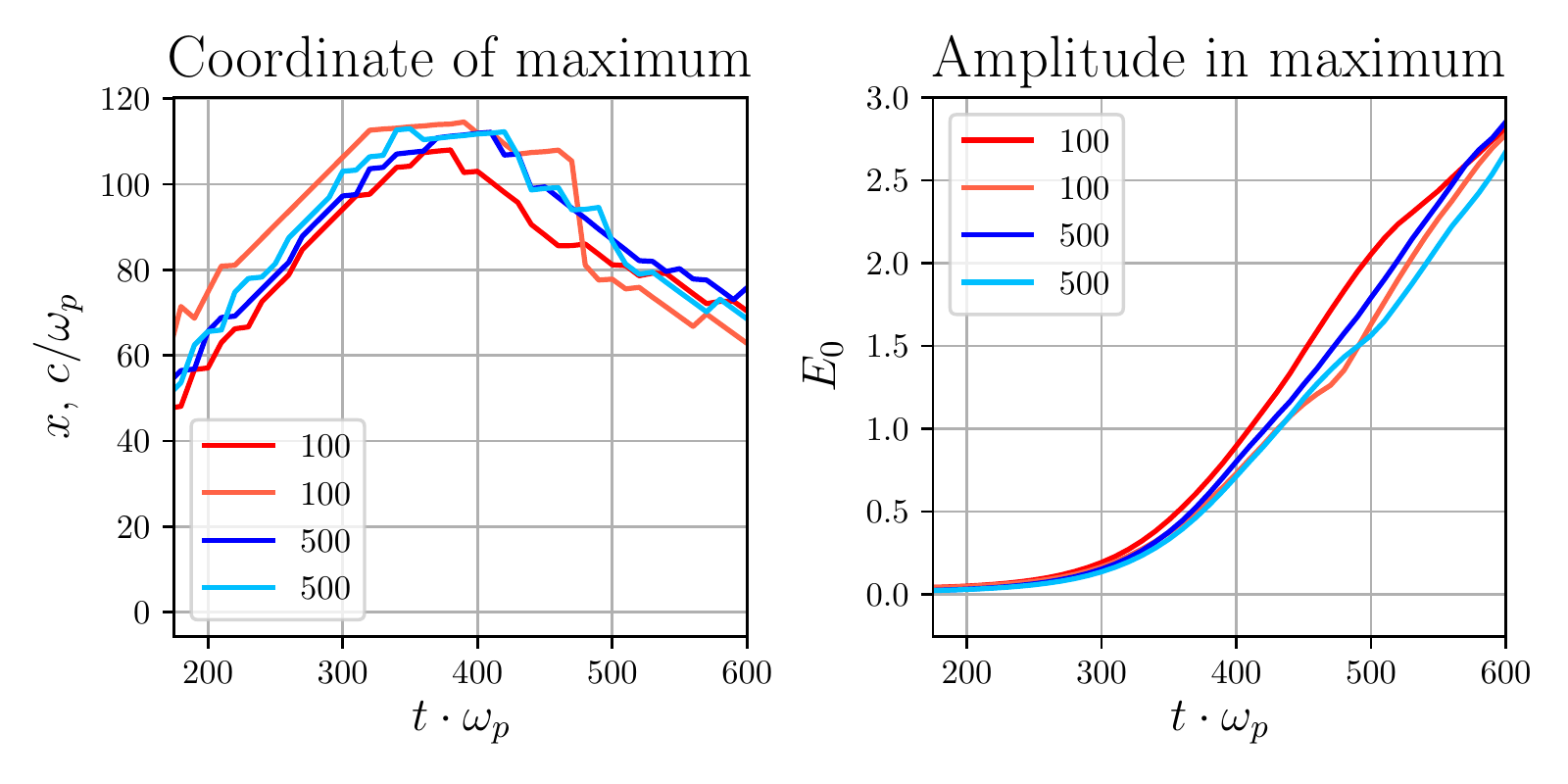}
	\caption{The $x$-coordinate of the maximal amplitude $E_0$ of plasma oscillations  and its absolute value in the case of a single beam.}
	\label{fig:MaxLoc}

\end{figure}

To estimate the location of the beam relaxation region, we use the amplitude of plasma oscillations averaged over the length of plasma oscillations $2l=2\pi c/\omega_p$ in the longitudinal direction and integrated over the plasma thickness $L_y$ in the transverse direction:
\begin{gather}
E_0(x,t)=\int\limits_{0}^{L_y}dy\left[\dfrac{\omega_p}{c\pi}\int\limits_{-l}^{l}dx'E_x^2(x+x',y,t)\right]^{1/2}.
\end{gather}
After calculation of $E_0(x,t)$ at each diagnostic time step, the value of the amplitude maximum as well as its longitudinal position are determined. Figure \ref{fig:MaxLoc} shows the results obtained by injection of a single beam into a plasma with fixed ions  in the found efficient regime using different numbers of $N_p$ particles in a cell. Shades of red denotes runs with $N_p = 100$, blue ones corresponds to $N_p = 500$.

\begin{deluxetable*}{lllcccccc}
	\tablecaption{Parameters of all two-beams  simulations presented in the article. \label{table:param}}
	\tablehead{\colhead{}&\colhead{$\Omega_e/\omega_p$}&\colhead{case name}&\colhead{$m_i/m_e$}&\colhead{$L_x{\times}L_y,$ $c/\omega_p$}&\colhead{$n_b/n_0$}&\colhead{$T^{(e)},$ eV}&\colhead{$T^{(b)}_\parallel/T^{(b)}_\perp,$ keV}&\colhead{$\sigma^{(b)},$ $c/\omega_p$}}
	\startdata
	run 1 - run 5  & 0.29&  \textit{Efficient regime}&$\infty$& $116.8\times32$& \multirow{5}{*}{0.005}&\multirow{5}{*}{80} & \multirow{5}{*}{$207.36/25.0801$} &\multirow{5}{*}{20}\\
	run 6 - run 10 & 0.15&  \textit{Weak field}&$\infty$&$116.8\times32$& & &  & \\
	run 11 - run 13 & 0&  \textit{Unmagnetized}&$\infty$& $108.8\times32$& & &  & \\
	run 14 - run 16 & 0.75&  \textit{Strong field}&$\infty$ &$156.8\times32$& & & &\\
	run 17 - run 19 & 0.29&  \textit{Eff. regime + ions}&$1836$& $116.8\times32$& & & & \\
	\enddata
\end{deluxetable*}

It is seen that, after the initial excitation of plasma waves at a sufficiently large distance from the injection point, a gradual shift of the relaxation region towards the injector is observed subsequently. The amplitude of oscillations grows continuously. In the absence of significant ion dynamics (fixed or very heavy ions, weak instabilities), this behavior is typical. Otherwise, the ponderomotive force of spatially localized wave packets and modulation instability result in formation of density perturbations. The presence of sharp density gradients in this area leads to a local breakdown of the beam instability. These effects have been considered in more details in \cite{Annenkov2019} where the injection of a thin sub-relativistic beam into a magnetized plasma is investigated.

Figure \ref{fig:MaxLoc} shows that dynamics of both the position and amplitude of the localized wave packet does not strongly depend on the number of particles, that is why we find it optimal to use 100 particles in a cell. This choice allows to collect statistics on the basis of a series of calculations for each considered set of system parameters using the available computing resources.

The plasma length in the efficient regime is chosen to be $L_x = 116.8$ $c/\omega_p$ ($2920$ cells). The characteristic total number of model particles is $0.6$ billion. There is no significant  EM emission during the injection of a single beam into the plasma.

We should note that the approach used here to determine the location of the beam relaxation region based on simulations of a single beam is approximate. In the case of  two beams injection, the conditions for the instability development changes. Therefore, the relaxation distances also become different. A larger relative density of the beams should lead to larger differences. A huge number of calculations with different distances between the injectors of the beams should be carried out  in order to determine the ideal overlapping. In addition, it is necessary to carry out a series of calculations for each plasma length, since the results of the beam instability in several simulations with different certain distribution functions can be slightly different. Such optimization requires a significant amount of computing resources and is redundant to answer the main questions of this article.

\section{Two beams simulations}\label{SimRes}

First, let us consider the case of beams injection into a plasma with immobile (or infinitely heavy) ions. The \textit{efficient regime} of emission with a magnetic field $\Omega_e/\omega_p = 0.29$ found in Sec. \ref{LinEffReg} will be compared with ones in which the maximal growth rate of the beam-plasma instability does not differ significantly in absolute value, but is located outside the three-wave interaction region. To go into these regimes, we will vary only the magnitude of the external magnetic field. We consider the cases of \textit{weak field} ($\Omega_e/\omega_p = 0.15$), \textit{strong field} ($\Omega_e/\omega_p = 0.75$) as well as the case of \textit{unmagnetized} plasma ($\Omega_e/\omega_p = 0$). For each case, we will carry out several (from three to five) runs with identical initial macro-parameters, but with a different specific implementation of the particle distribution function. We will use the ratio of the radiation power to the power introduced into the system by both beams as a comparative parameter for radiation efficiency. Then we consider the effect of ion dynamics for the efficient regime on the example of hydrogen ions with the real mass  $m_i/m_e=1836$. The parameters of all calculations are given in the table \ref{table:param}.

\subsection{Immobile ions}

\begin{figure*}[h!]
	\centering
	\includegraphics[width=0.99\linewidth]{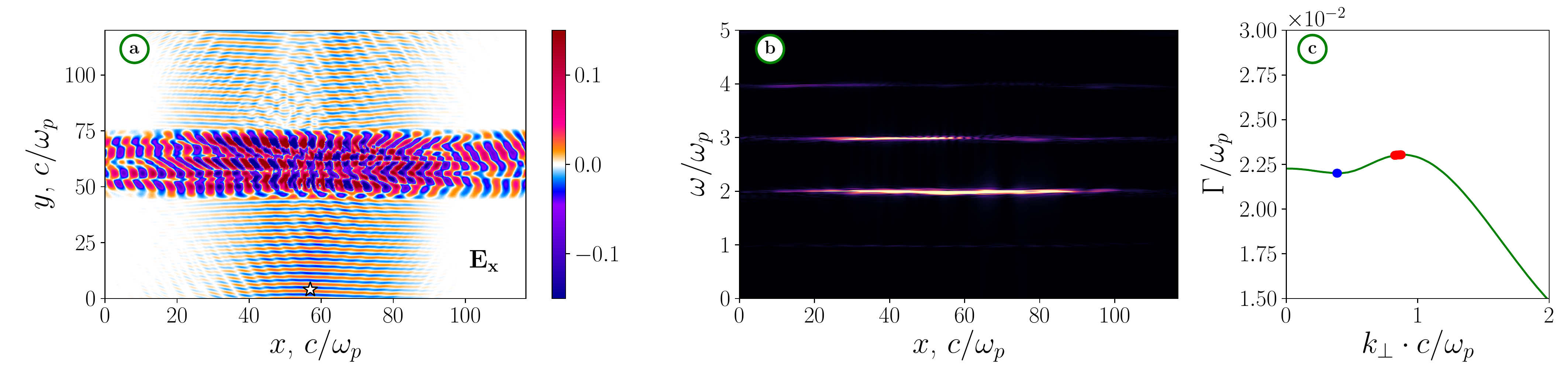}
	\caption{\textbf{Efficient regime.} (a) The map of electric field $E_x$ in the moment $t=728\omega_p^{-1}$; (b) radiation spectra of electric field components for the entire simulation time at the outer boundary of the vacuum region as functions of the longitudinal coordinate $x$; (c) theoretical growth rate $\Gamma(k_\perp)$ along the line of the local maximum.\label{PICeff}}
\end{figure*}

\begin{figure*}[h!]
	\centering	
	\includegraphics[width=0.99\linewidth]{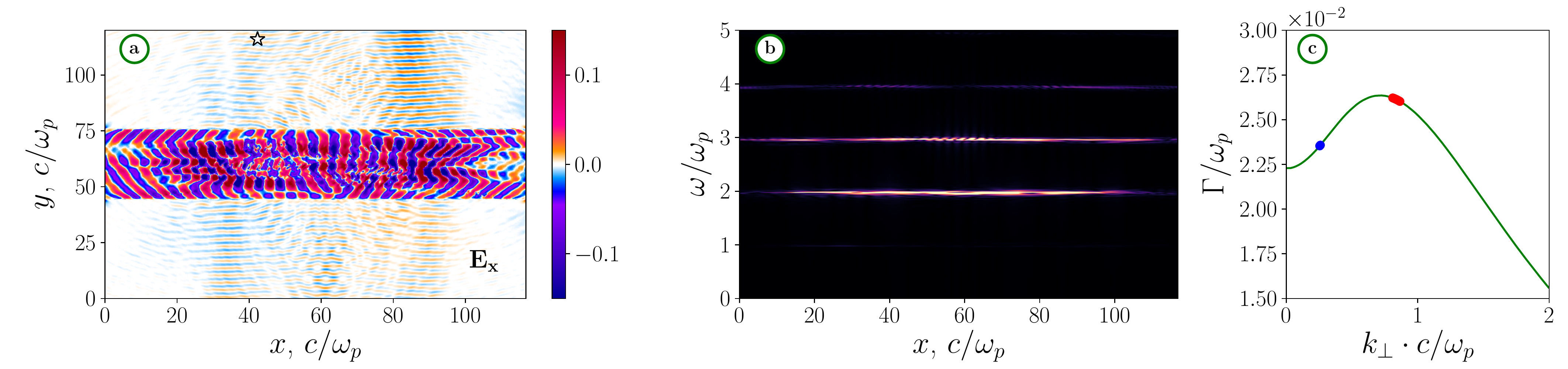}
	\caption{\textbf{Weak field.}  $t=728\omega_p^{-1}$.\label{PICweak}}
\end{figure*}

\begin{figure*}[h!]
	\centering	
	\includegraphics[width=0.99\linewidth]{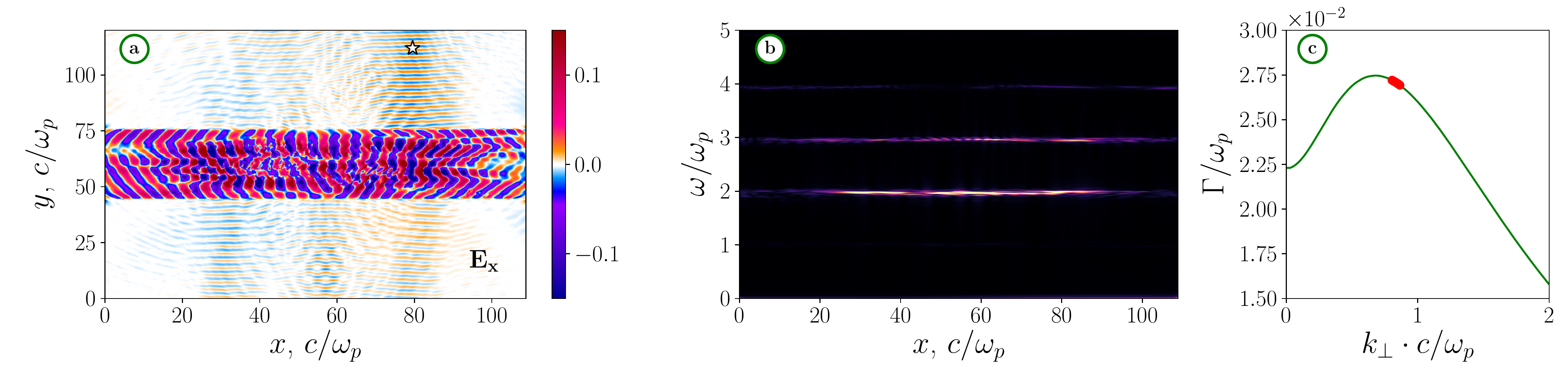}
	\caption{\textbf{Unmagnetized.}  $t=848\omega_p^{-1}$.\label{PICunmag}}
\end{figure*}

\begin{figure*}[t]
	\begin{center}
		\includegraphics[width=1\linewidth]{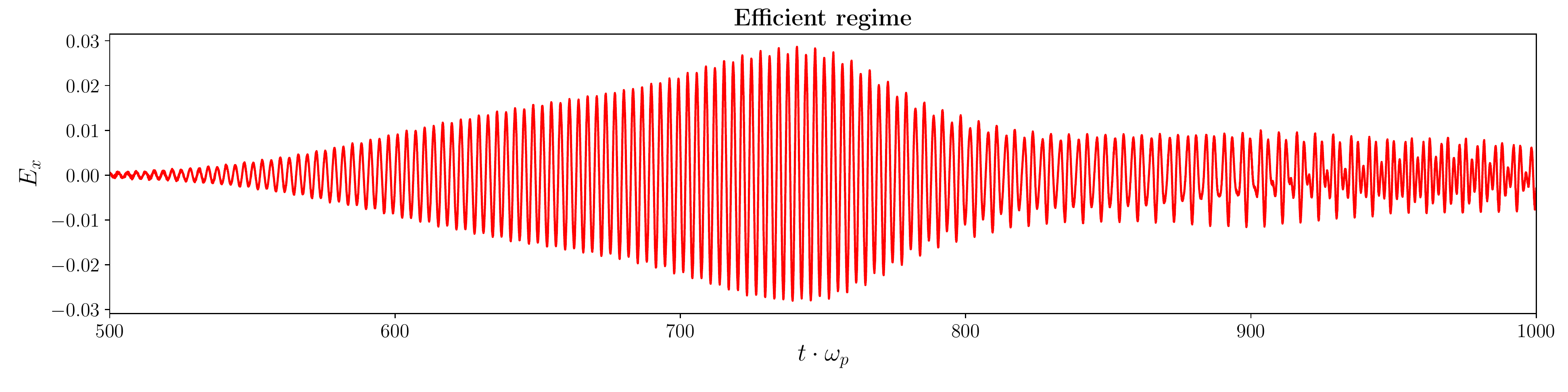}	
		\caption{The temporal dependence of the $E_x$-field in the single point indicated by the white star in Figure \ref{PICeff}.\label{fig:1dfield}}
		\begin{minipage}[h]{0.24\linewidth}
			\includegraphics[width=1\linewidth]{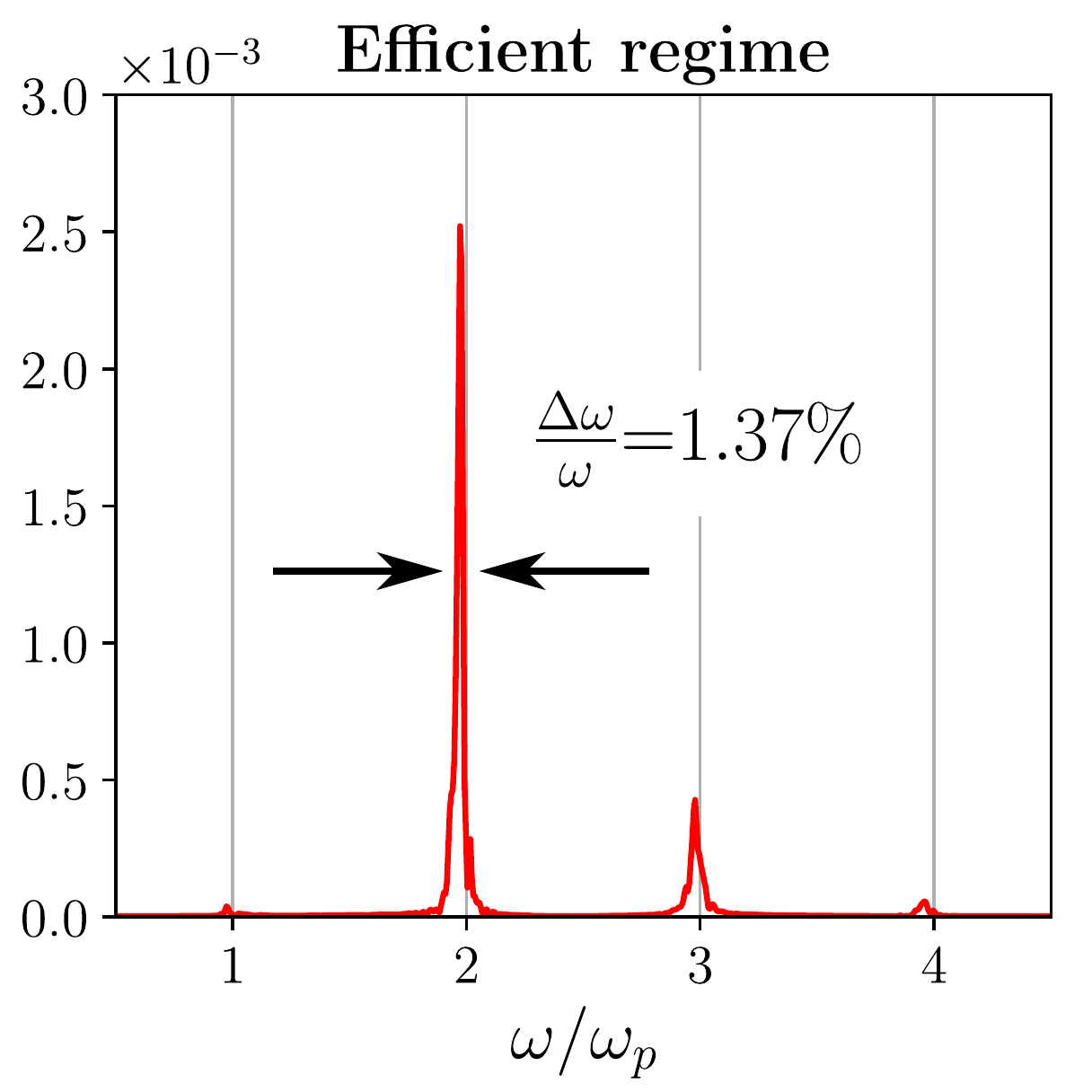}
		\end{minipage}
		\hfill
		\begin{minipage}[h]{0.24\linewidth}
			\includegraphics[width=1\linewidth]{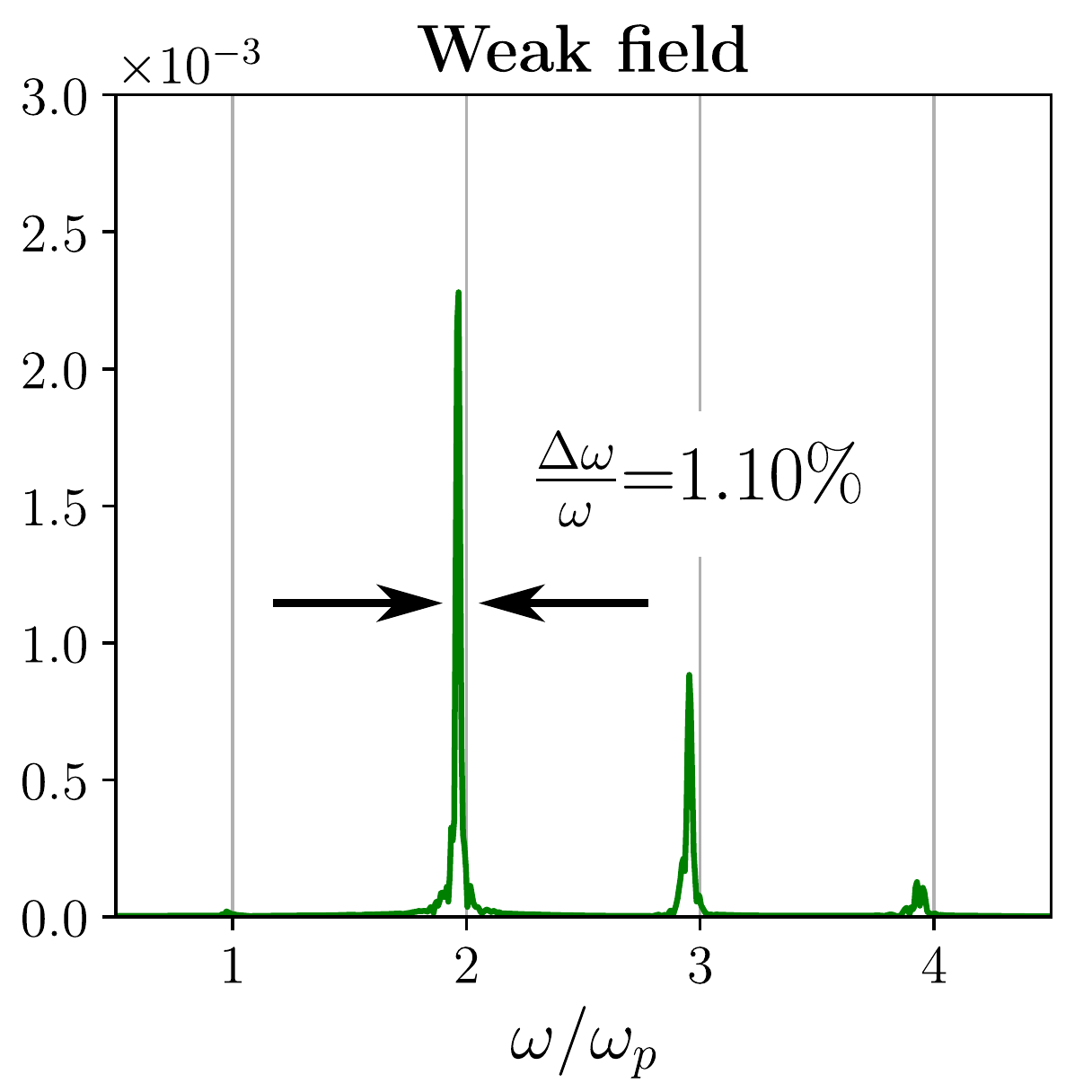}
		\end{minipage}
		\hfill
		\begin{minipage}[h]{0.24\linewidth}
			\includegraphics[width=1\linewidth]{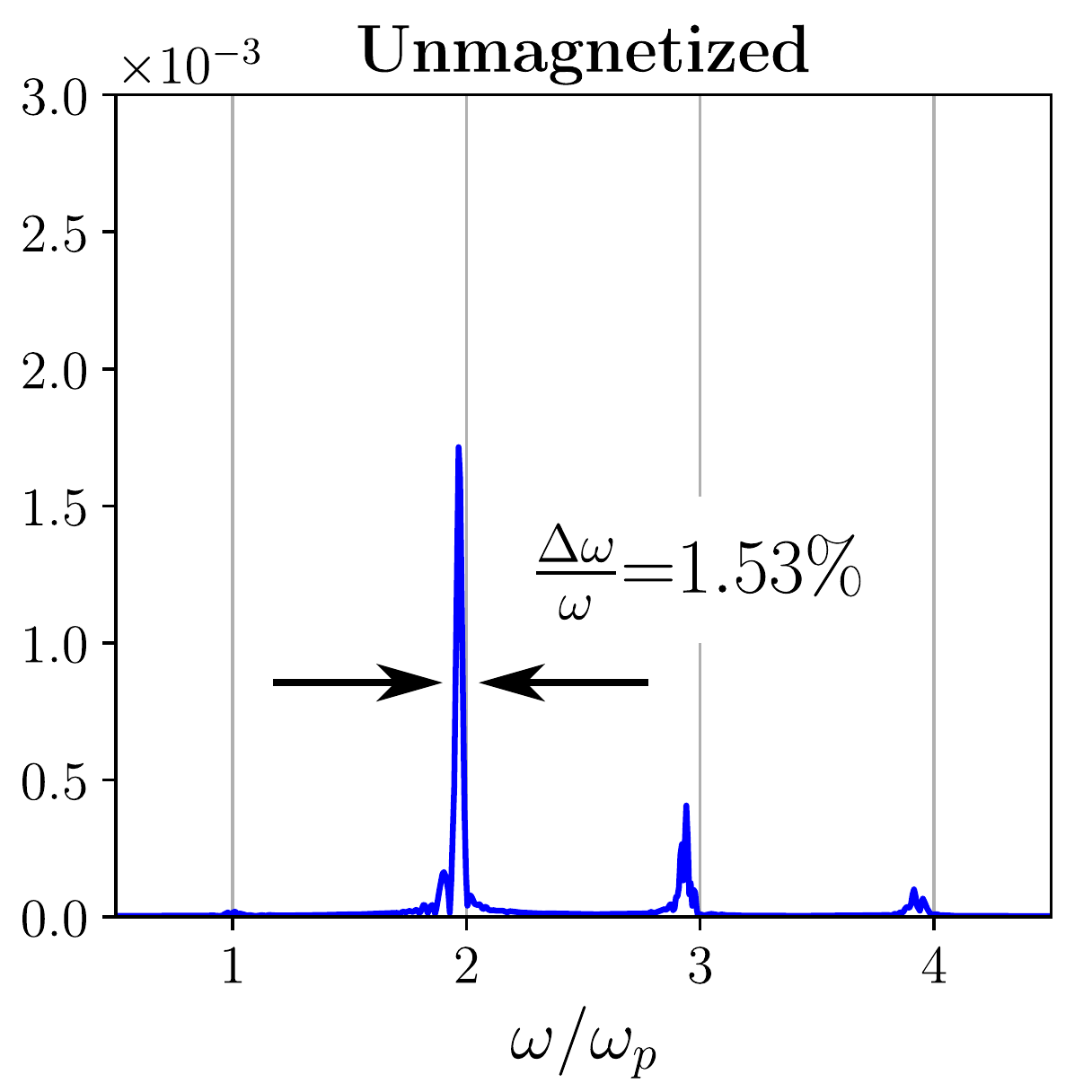}
		\end{minipage}
		\hfill
		\begin{minipage}[h]{0.24\linewidth}
			\includegraphics[width=1\linewidth]{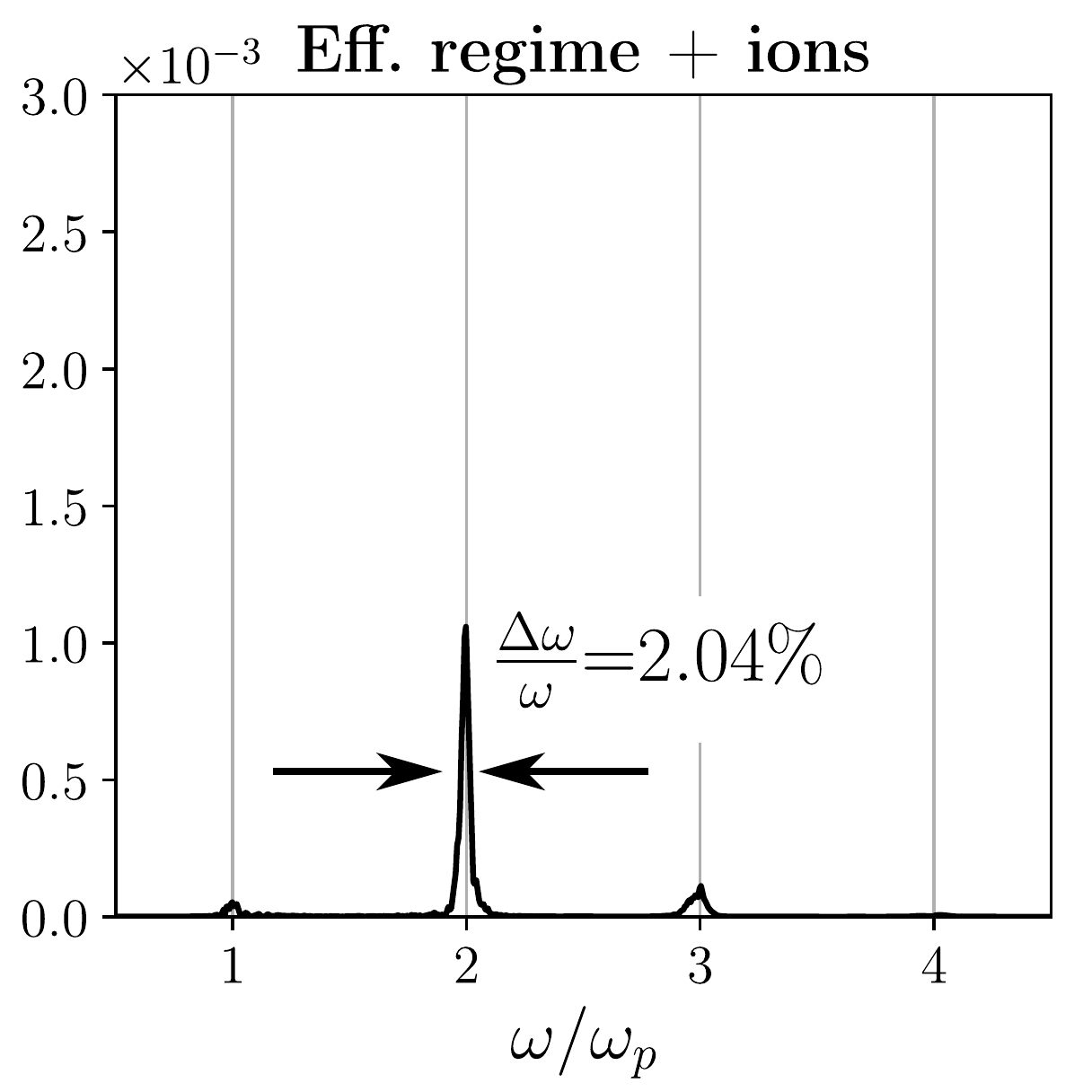}
		\end{minipage}
		\caption{The frequency spectrum of the produced radiation in the single point indicated by the white star on corresponding Figures \ref{PICeff},\ref{PICweak},\ref{PICunmag} and \ref{PICions}.\label{fig:FFT}}	
	\end{center}
\end{figure*}

Figures  \ref{PICeff},\ref{PICweak} and \ref{PICunmag}  show the results of PIC simulations corresponding to the most efficient realizations of each mentioned regimes with immobile ions. In subplot (a) we present the maps of the longitudinal field $E_x$ at the moment of radiation maximum, in subplot (b) -- emission spectra for the entire simulation time recorded along the boundary of the computational domain, in (c) -- graphs of the instability growth rate  in the framework of the exact linear theory. It is seen that weakening the magnetic field to zero  increases slightly the maximal value of the growth rate and removes it away from the three-wave interaction region in $k$-space. A decrease in the level of electromagnetic emission is evident from the electric field maps (subplot (a)). It follows from subplots (b) that the main spectral characteristics of the radiation are the same in all regimes: the radiation near the doubled plasma frequency dominates and there is also a noticeable peak near the third harmonic. Figure \ref{fig:1dfield} presents the temporal dependence of the radiation field $E_x$ in $m_e\omega_p c/e$ units, recorded at one point on the boundary of the computational domain for run 1. The emission spectra recorded at the points indicated by the star in Figures \ref {PICeff}, \ref{PICweak} and \ref{PICunmag} are shown in Figure \ref{fig:FFT}. The emission line width near the second harmonic of the plasma frequency for all regimes is less than or of the order of 2\%.

\begin{figure*}[h!]
	\centering	
	\includegraphics[width=0.99\linewidth]{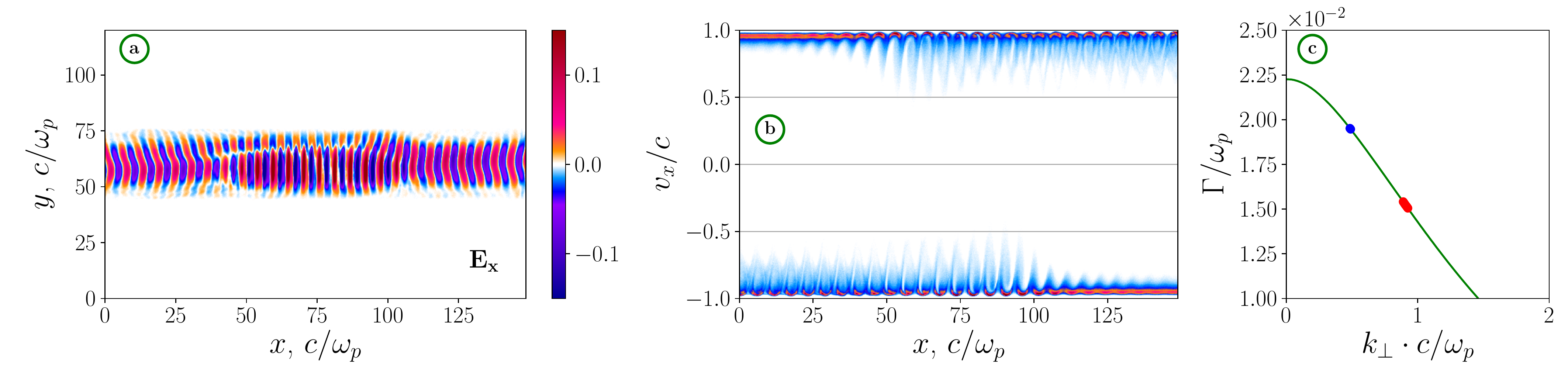}
	\caption{\textbf{Strong field.}  $t=728\omega_p^{-1}$. (b) phase space $(x,v_x)$ of the beam.\label{PICstr}}
\end{figure*}

The beam instability becomes purely longitudinal in the strong magnetic field (Figure \ref{PICstr}). In this regime, we also observe a noticeable extension of the beam relaxation length.  Therefore, we have extended the total system length in order to overlap beams relaxation regions more efficiently. The overlapping degree can be estimated from the image of the phase space $(x, v_x)$ (Figure \ref{PICstr} (b)). Figure \ref {PICstr} (a) shows that, in the case of a sufficiently strong external magnetic field, electromagnetic emission disappears.

\begin{figure}[h!]
	\centering
	\includegraphics[width=1\linewidth]{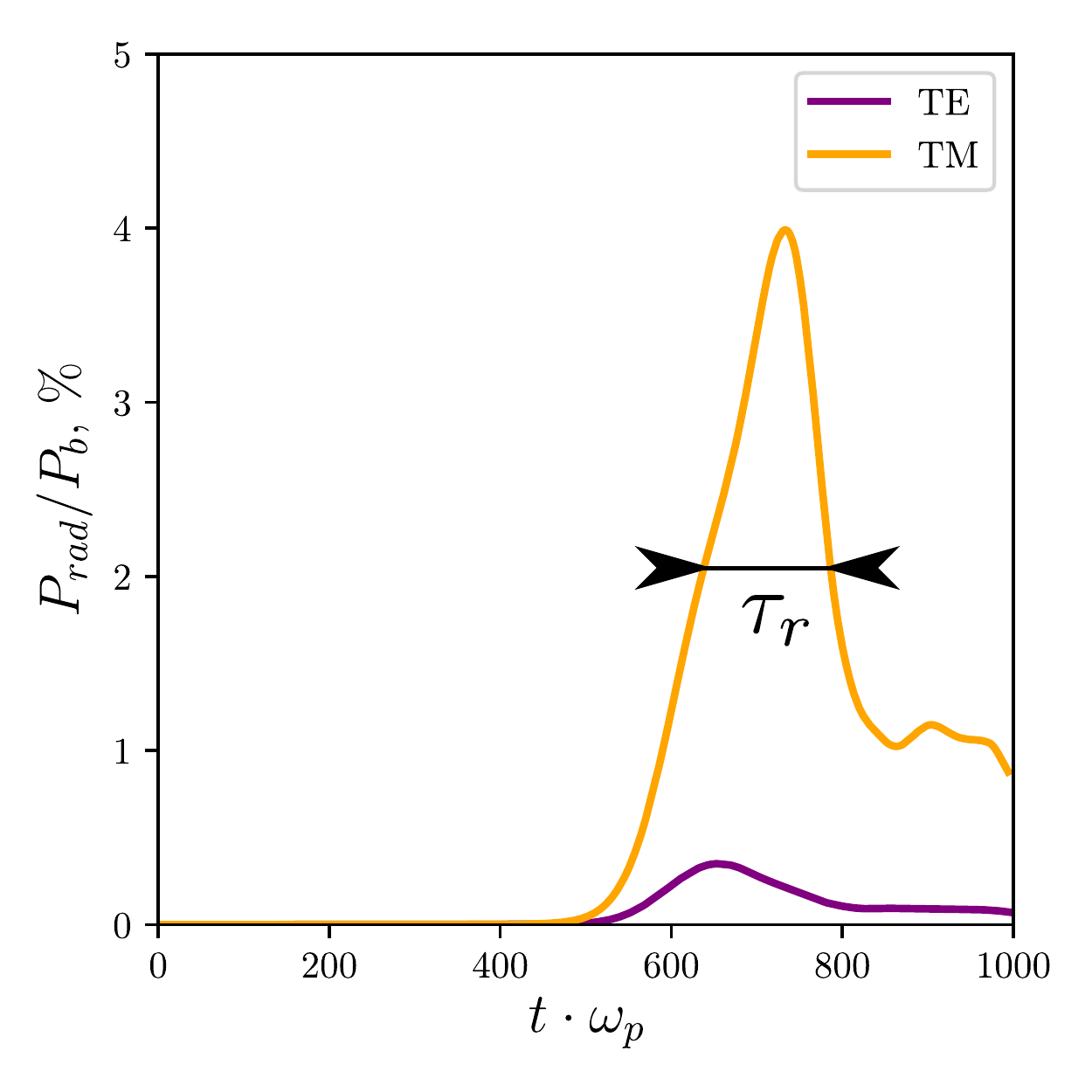}
	\caption{The efficiency of beam-to-radiation power conversion as a function of time for run 1.\label{fig:eff}}
\end{figure}

Let us consider the efficiency of EM radiation generation in these regimes quantitatively. Figure \ref{fig:eff} presents the fraction of the total beam power $P_b$ converted into the radiation power $P_{rad}$ for run 1 as a fuction of time (\textit{efficient regime}). In this graph, we separate contributions from various radiation polarizations (TM {\bf or O} mode includes the fields $E_x$, $E_z$ and $B_y$, TE {\bf or X} mode -- $B_x$, $B_z$, $E_y$). It is seen that the ordinary TM mode dominates and the efficiency of its generation reaches 4\% of the beam power. The total beams-to-radiation energy conversion efficiency  is $\approx10^{-2}$. The duration of the main emission can be estimated as a FWHM width of the plot shown in Figure \ref{fig:eff}. In all simulations performed for the efficient regime, this duration lies in the range $\tau_r\approx160\div220\omega_p^{-1}$. 

\begin{figure*}[t]
	\begin{center}
		\includegraphics[width=0.99\linewidth]{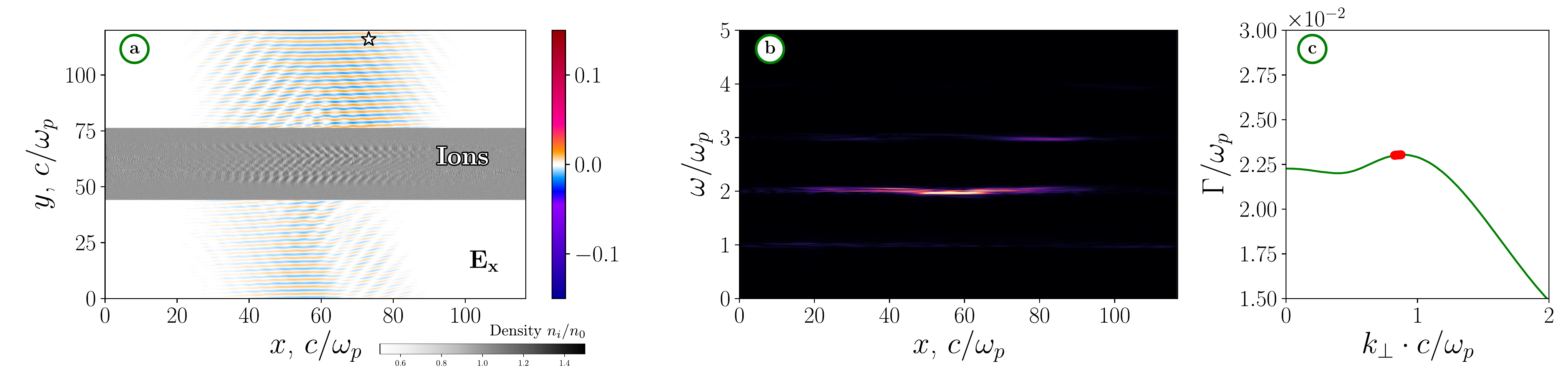}
		\caption{\textbf{Eff. regime + ions.}  $t=560\omega_p^{-1}$. (a) Ion density in plasma column and radiation field $E_x$ in vacuum. (b) Frequency spectrum of radiation field along the boundary of simulation box. (c) Theoretical growth rate along the line of its local maximum.}\label{PICions}	
	\end{center}
\end{figure*}

\begin{figure}[H]
	\includegraphics[width=0.99\linewidth]{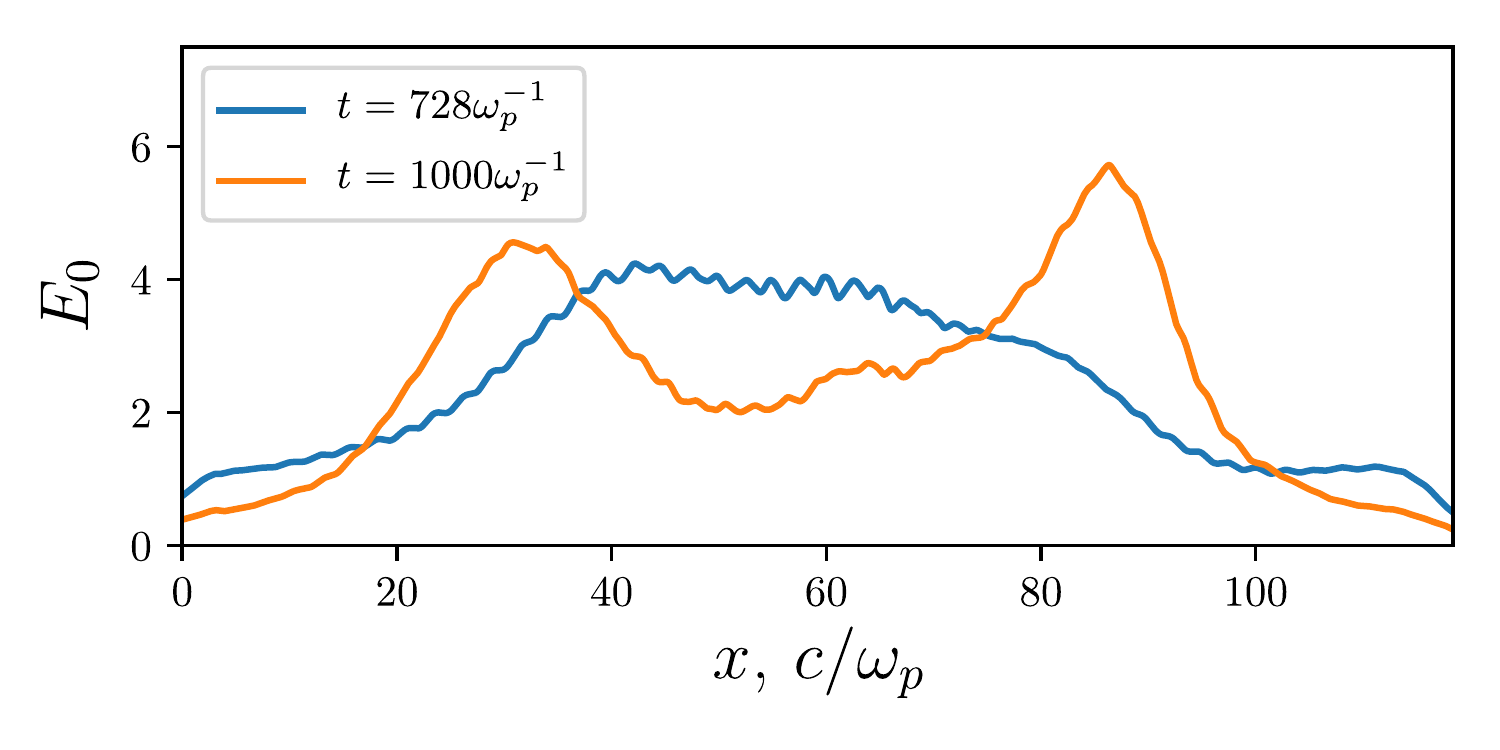}
	\caption{$E_0(x)$ at two time moments for run 1.\label{fig:E0TwoBeams}}
\end{figure}

The radiation efficiency depends not only on the amplitudes of the plasma oscillations excited by the beams, but also on the degree of their overlapping. For this reason, the dependence of the emission level on time has a pronounced maximum corresponding to the moment when the beams excite counterpropagating plasma waves not only with sufficiently large amplitudes, but also with localization in the same region. At later times, the beams continue to excite plasma oscillations, but the radiation efficiency decreases due to the shift of their relaxation regions. This fact is confirmed by Figure \ref{fig:E0TwoBeams} showing the dependence of the plasma wave amplitude $E_0(x,t)$ on the longitudinal coordinate in two time-moments when  the emission efficiency reaches its maximum ($t = 728 \omega_p^{-1} $) and  when it is significantly decreased ($t=1000\omega_p^{-1}$).

\begin{figure}[h!]
	\centering
	\center{\includegraphics[width=1\linewidth]{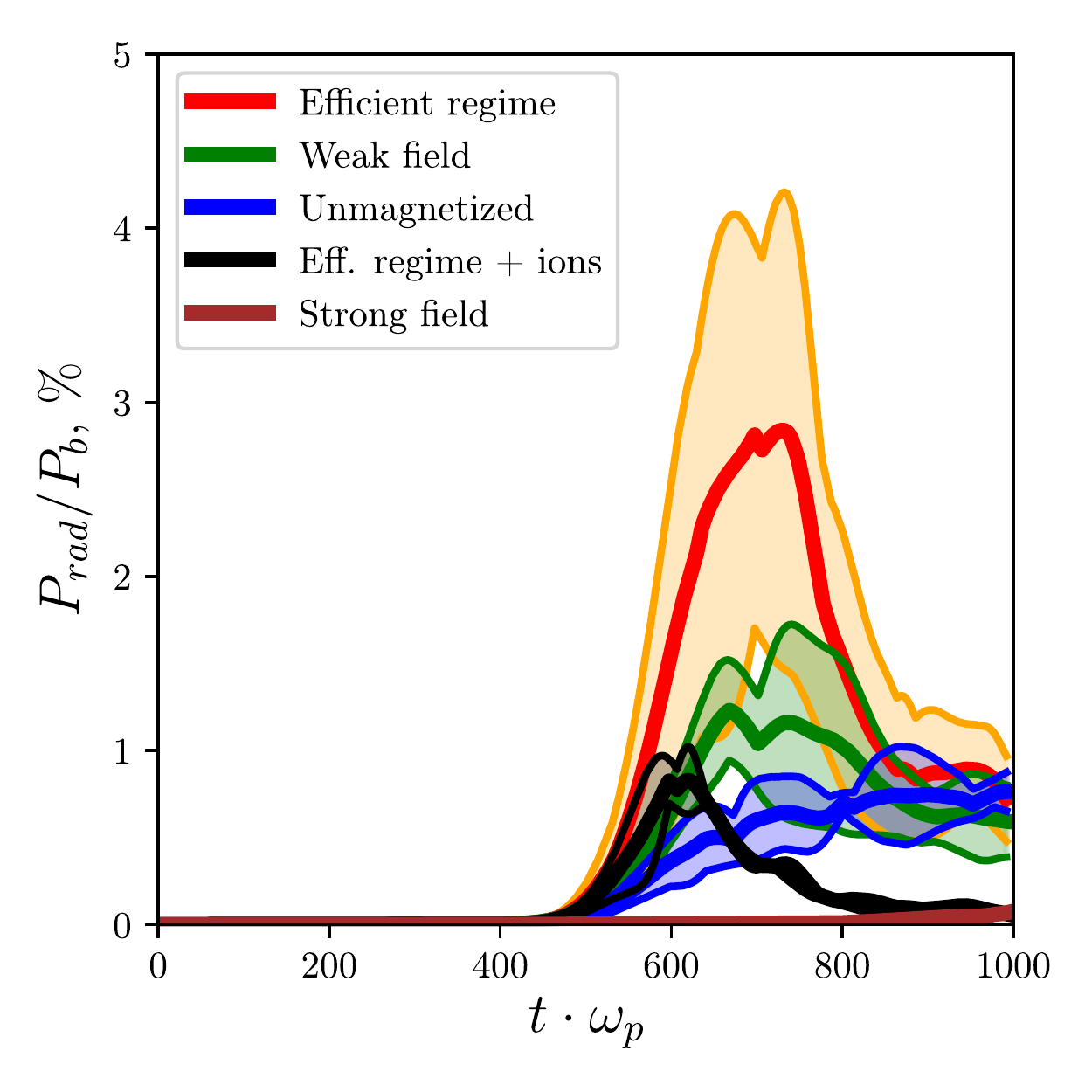}}
	\caption{The efficiency of beam-to-radiation power conversion as a function of time for all runs.\label{fig:eff2}}
	
\end{figure}
Figure \ref{fig:eff2} shows the graphs of the total efficiency (TM + TE) for all simulations with colliding beams performed in this work. The methodology for constructing these dependencies is as follows. In each moment of time, from all the runs corresponding to one regime, the highest and lowest radiation efficiencies are found and placed on the graph. The thick line indicates the average value between them. Despite the significant scatter of results for the \textit{efficient regime}, the typical value of the generation efficiency is several percent. When switching to regimes with lower magnetic fields, the efficiency decreases several times. In the case of a strong field, there is no noticeable radiation.

\subsection{Effect of ion dynamics}

Accounting for the ion dynamics leads to the formation of density modulation (Figure \ref{PICions} (a)) which, in turn, leads to a local breakdown of the two-stream instability and shifts the region of intense beam relaxation away from the injector. As a result, substantial emission turns out to be possible only at the initial stage of the beam-plasma interaction. Therefore, the final generation efficiency reaches only a value of the order of one percent. The duration of the radiation is reduced to a value of the order of $\tau_r^{(i)}\approx150 \omega_p^{-1} $. In this case, the radiation spectrum (Figure \ref{fig:FFT} and \ref{PICions} (b)) does also undergo some changes. Emissions near the third and fourth harmonics are significantly weakened, since the nonlinear processes responsible for such emissions arise at the late stage of the beam-plasma interaction that is strongly affected by the formed density modulation. Despite the general decrease in the level of electromagnetic emission in the case of mobile ions, there is some increase in the fraction of radiation near the plasma frequency (Figure \ref{fig:FFT}), which is apparently associated with the beam-plasma antenna mechanism \citep{Timofeev2015, Annenkov2016, Annenkov2016b, Timofeev2016a}.

In these simulations, we assume hydrogen plasma. Obviously, the effect of ion dynamics will decrease upon transition to heavier atoms. In case of sufficiently heavy gases, the radiation generation efficiency and its spectrum should not differ significantly from the case of immobile ions.

\subsection{Results in dimensional units}

Let us evaluate the results in dimensional units. The power of each beam under the assumption of their axial symmetry with diameter $\sigma^{(b)}=20$~$c/\omega_p$ is
\begin{gather}
P_b=(\gamma-1) \dfrac{n_b}{n_0} \dfrac{v_b}{c} \pi \left(\dfrac{\sigma^{(b)}\omega_p}{2c}\right)^2 P_0\approx 2\, \text{GW}.
\end{gather}
Here $P_0=m_e^2 c^5/(4\pi e^2)=0.69$~GW. Thus, the total power of the injected beams is $\approx4$~GW. Therefore the maximum emission power corresponding to the efficiency $4$\% reaches the value $\approx160$~MW.

\begin{figure}[h!]
	\includegraphics[width=1\linewidth]{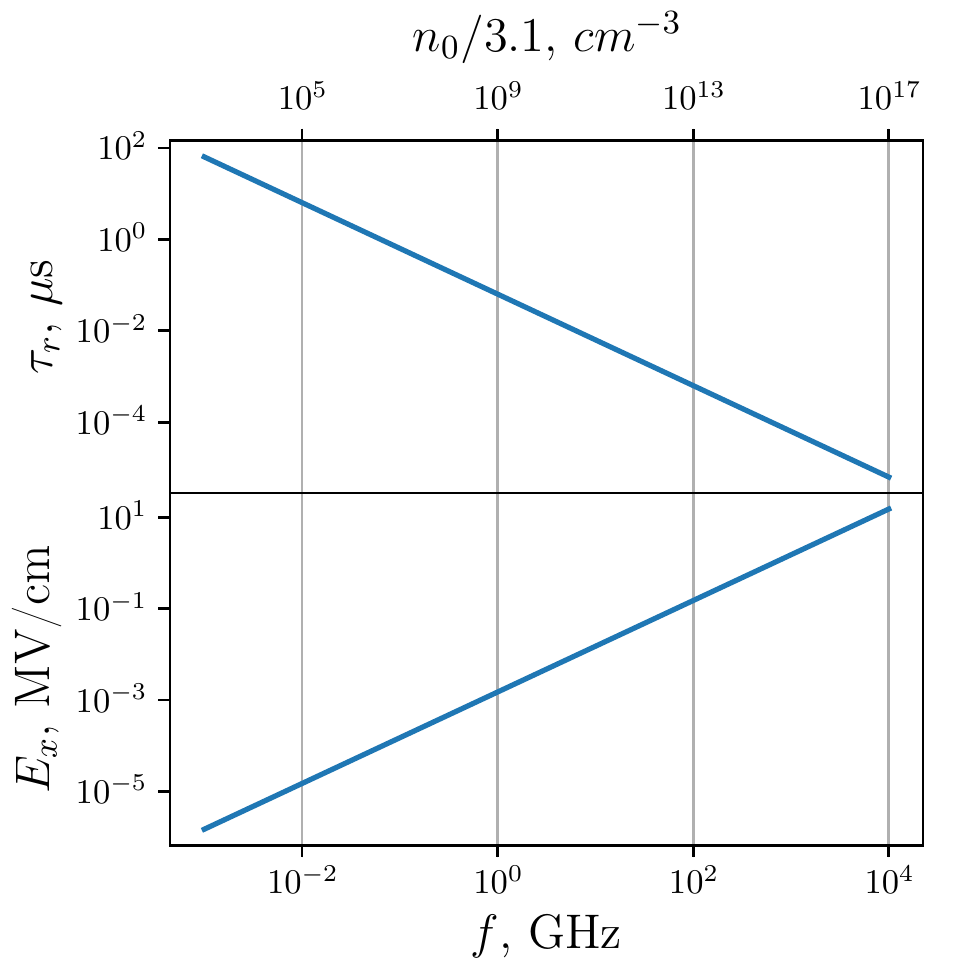}
	\caption{Radiation duration $\tau_r=200\omega_p^{-1}$ and maximum field amplitude in dimensional units for different emission frequencies.\label{fig:dim}}	
\end{figure}

In our simulations, we use dimensionless units depending on the plasma frequency $\omega_p=\sqrt{4\pi e^2 n_0/m_e}$ where $n_0$ is the plasma density. Since the emission frequency is tied to the plasma frequency ($f = 2\omega_p/2\pi$), by setting the desired radiation frequency, we obtain the corresponding plasma density $n_0$ and also determine all other parameters of the system. Figure \ref{fig:dim} presents the dependences of the radiation pulse duration $\tau_r$ in ns and the maximum amplitude of the radiation field in MV/cm on the radiation frequency $f$ and the corresponding plasma density $n_0$. In particular, it means that, in order to demonstrate the existence of efficient radiation regime  in laboratory experiments at the frequency $f = 1$ THz, one should create a plasma with the electron density $n_0 = 3.1 \cdot10^{15}$ cm$^{-3}$ and length $L_x = 1.1$ cm. Counter-injection into this plasma of axially symmetric electron beams with the diameter $\sigma^{(b)} = 2$  mm and current density $j\approx 70$  kA/cm$^2$ will produce a pulse of the second harmonic EM emission with the typical duration $\tau_r = 60 $ ps and electric field amplitude $1.5$ MV/cm.

\section{Discussion \& Conclusion}\label{Conc}

In this work, we have studied the regime of enhanced second harmonic EM emission from a plasma with colliding electron beams. Using the exact linear theory of the beam-plasma instability for the given relative beam density $n_b/n_0=0.005$ and energy $E_b\approx1$ MeV, we have found a regime in which the most unstable counterpropagating beam-driven modes satisfy the condition of three-wave interaction with an electromagnetic wave. The feasibility of this regime is confirmed via PIC simulations in a realistic model with continuous injection of electron beams into a finite-size plasma. The maximal conversion efficiency of the beam power into the emission power is found to reach the value $4$\%. We have also considered regimes with slightly differing system parameters, in which the maximum of the instability growth rate comes out from the region of three-wave interaction, but does not significantly change its absolute value. A decrease in the radiation efficiency by several times is observed in these regimes. This result, as well as a significant scatter in values of the radiation efficiency from shot to shot, allow to conclude that the EM emission level is strongly sensitive to the accuracy of hitting the growth rate maximum in the region of the discussed three-wave interaction.

Significant generation of electromagnetic radiation is not observed in the case of strong enough longitudinal magnetic field capable of suppressing the dominance of oblique instabilities. Accounting for the ion dynamics in the hydrogen plasma leads to the formation of density modulation. It results in a local disruption of the beam instability and a decrease of the EM radiation level.

Simulations with open boundary conditions presented in this paper allow to study EM emission processes in a beam-plasma system at a quantitatively new level. In particular, it becomes possible to imagine how the discussed three-wave process will proceed in a laboratory experiment. We have shown that the feasibility of efficient three-wave interaction regime can be proved experimentally at THz frequencies by injecting a pair of weakly relativistic kiloampere electron beams with nanosecond durations into a cm-scale gas jet. The same three-wave process at lower magnetic fields and beams densities can contribute to solar emissions (for example, type II radio bursts) allowing the presence of counterstreaming flows of accelerated electrons.   

\section*{Acknowledgments}
Simulations were performed using the computing resources of the Center for Scientific IT-services ICT SB RAS (\href{https://sits.ict.sc}{https://sits.ict.sc}) and ''Govorun'' supercomputer in JINR (\href{http://hlit.jinr.ru/}{http://hlit.jinr.ru/}).

This work was supported by RFBR (Grant No. 18-02-00232).

\software{CUDA \citep{cuda},
openMPI \citep{open_mpi}, Matplotlib \citep{matplotlib}, Inkscape \citep{Inkscape}.}

\bibliography{references} 

\end{document}